\documentstyle[manuscript,eqsecnum,aps,epsfig]{revtex}
\begin{document}
\title{
The Emission Pattern of High-Energy Pions: A New Probe for the 
Early Phase of Heavy Ion Collisions}
\author{A.\,Wagner
        \footnote[1]{Present address: Forschungszentrum Rossendorf, 
                     D-01314\,Dresden, Germany},
        C.\,M{\"u}ntz
        \footnote[2]{Present address: Universit{\"a}t Frankfurt, 
                     D-60054\,Frankfurt, Germany},
        H.\,Oeschler,
        C.\,Sturm}
\address{Technische Universit{\"a}t Darmstadt, D-64289\,Darmstadt, Germany}
\author{R.\,Barth,
        M.\,Cie\'slak,
        M.\,D\c{e}bowski$^*$,
        E.\,Grosse$^*$,
        P.\,Koczo{\'n},
        F.\,Laue
        \footnote[3]{Present address: Ohio State University, 
                     Columbus, OH-43210, USA},
        M.\,Mang,
        D.\,Mi{\'s}kowiec,
        E.\,Schwab,
        P.\,Senger}
\address{Gesellschaft f{\"u}r Schwerionenforschung, 
         D-64291\,Darmstadt, Germany}
\author{P.\,Beckerle,
        D.\,Brill,
        Y.\,Shin,
        H.\,Str{\"o}bele}
\address{Johann Wolfgang Goethe Universit{\"a}t, 
         D-60054\,Frankfurt, Germany}
\author{W.\,Walu{\'s}}
\address{Uniwersytet Jagiello{\'n}ski, 
         PL-30-059\,Krak{\'o}w, Poland}
\author{B.\,Kohlmeyer,
        F.\,P{\"u}hlhofer,
        J.\,Speer,
        I.K.\,Yoo}
\address{Phillips-Universit{\"a}t Marburg, 
         D-35037\,Marburg/Lahn, Germany}
\maketitle
\begin{abstract}
The emission pattern of charged pions has been measured 
in Au+Au collisions at 1\,GeV/nucleon incident energy.
In peripheral collisions and at target rapidities, high-energy pions 
are emitted preferentially towards the target spectator matter.
In contrast, low-energy pions are emitted predominantly 
in the opposite direction. 
The corresponding azimuthal anisotropy is explained by the interaction 
of pions with projectile and target spectator matter. 
This interaction with the spectator matter causes an effective 
shadowing which varies with time during the reaction.
Our observations show that high-energy pions stem from the early stage 
of the collision whereas low-energy pions freeze out later.
\end{abstract}

Heavy-ion collisions at relativistic energies provide a unique possibility
to study nuclear matter at high densities and at high temperatures in 
the laboratory.
These reactions last only several $10^{-23}$\,s and
within this time interval the baryonic density varies between about three 
times normal nuclear matter density ($\rho_0$ = 0.17 \rm{fm}$^{-3}$) in the 
early phase and about $\rm{0.2} \times \rho_0$ at freeze out when the 
particles cease to interact~\cite{Stock:1986,Aichelin:1991}.
A prerequisite for the study of the properties (e.g. the nuclear equation 
of state) of the dense ``fireball'' is to obtain information on the  
space-time evolution of the nuclear matter distribution in the course of 
the collision.

Pions are considered to be a sensitive probe of the reaction dynamics.
They are produced  abundantly  and due to the large $\pi N$   
cross section pions are continuously 
``trapped'' by forming baryonic resonances 
(e.g.~$\pi N \rightarrow \Delta$) 
which then can decay by pion emission. 
Therefore, pions - especially those with momenta between 
0.2\,GeV/c and 0.5\,GeV/c - are expected to freeze out predominantly  
in the late and dilute stage of the collision. High-energy pions, however,
interact less strongly with the nucleons and hence have a chance to
decouple already in an early phase.
Therefore, a detailed study of high-energy pions may shed light
on the hot and dense stage of the collision
as suggested in\cite{Muentz:1995}.
First evidence for different freeze-out conditions was extracted  
from the $\pi^-$/$\pi^+$ ratio 
measured in central Au+Au 
collisions at 1\,GeV/nucleon. The $\pi^-$/$\pi^+$ ratio as a function of 
transverse momentum was analyzed  in terms of Coulomb
interaction between pions and the nuclear fireball. 
It was found that high-energy pions are emitted from a more compact 
source than low-energy pions \cite{Wagner:1998}.

Our experimental approach to investigate the space-time evolution of the pion
source is to exploit the absorption or rescattering of pions when interacting 
with the spectator fragments.
The shadow cast by the spectator matter leads to a depletion of pions
according to the emission time of the pions and
the motion of the spectator matter.
Preferential emission of pions in the reaction plane 
was found in asymmetric collisions and
has been interpreted as an effect of shadowing by a large target 
nucleus\cite{Gosset:1989}.
In symmetric collision systems, a preferential emission perpendicular to the 
reaction plane has been observed both for charged~\cite{Brill:1993} and 
neutral~\cite{Venema:1993} pions and has been interpreted as 
absorption or rescattering effects in the spectator matter.
Recently, an enhanced in-plane emission of pions was observed in Au+Au 
collisions\cite{Kintner:1997} with more (positive) pions being emitted 
opposite to the target spectator fragments.
Since in a hydro-dynamical interpretation the preferential motion of 
nucleons and composite particles towards the spectator fragments is
described as ``flow'', the complementary effect observed for pions was called 
``antiflow''.
The ``antiflow'' of pions is found to be pronounced only in peripheral 
Au+Au collisions and vanishes for central collisions\cite{Kintner:1997}.

In this Letter we present data on pion production in Au+Au collisions
at a bombarding energy of 1\,GeV/nucleon as a function
of the pion azimuthal emission angle, the rapidity,
the transverse momentum, and the collision centrality.
With this detailed information one can monitor  the effect of
spectator shadowing on the pion emission pattern
at subsequent stages of the collision. 
The fast moving spectator matter represents an obstacle for the pions
emitted from the fireball. This introduces a time scale which allows to  
follow the evolution of the pion source  and to study
correlations of pion energy and freeze-out time.
Emphasis is put on the investigation of high-energy pions which are measured
with high statistics.

The experiments were performed with the Kaon Spectrometer~\cite{Senger:1993}
at the heavy-ion synchrotron SIS at GSI (Darmstadt)
which delivered a beam of $^{197}$Au$^{65+}$ impinging onto
a 1.93\,g/cm$^2$ Au target.
The spectrometer covers a momentum-dependent solid angle of
$\Omega = 15-35\,$msr and a momentum bite of $p_{max}/p_{min} \approx 2$ 
for a given magnetic field setting.
The measured laboratory momenta vary between 0.156\,GeV/$c$ and 1.5\,GeV/$c$
with data taken in four different magnetic field settings.
The particle trajectories and momenta are reconstructed 
using three multi-wire proportional chambers. The particle velocities
are determined with two time-of-flight arrays. Both measurements allow
to identify pions up to 1.5\,GeV/$c$.     
The collision centrality is determined by means of the charged-particle
multiplicity measured in a polar angle range between 12 and 48 degrees 
using a 84-fold segmented plastic-scintillator detector.
For our study we select  peripheral collisions ($65\pm 5$\% of the reaction
cross section) and the $14\pm 4$\% most central collisions. 
The reaction cross section (5.9$\pm$0.4 barn) 
was measured with a minimum bias trigger
which required a charged particle multiplicity of more than two
in the polar angle range given above. 

The determination of the reaction plane in every collision is based on the 
measurement of charged projectile spectator fragments detected between 
$0.5^o \leq \theta_{lab} \leq 5^o$ using a plastic scintillator wall 
of 380 modules positioned 7 m downstream from the target.
The orientation of the event plane is determined by the sum of transverse 
momenta of the charged projectile spectator particles 
\cite{Danielewicz:1985,Brill:1997}.
The dispersion of the reaction plane amounts to 
$\approx 45^o$ for peripheral collisions\cite{Brill:1996}.

Figure~\ref{figure-1} depicts the nuclear matter distribution
for a Au+Au collision at a beam energy of 1\,GeV/nucleon at 
6.5, 12.5 and 16.5\,fm/$c$ after time zero (which is the time instant 
when both nuclei have a distance projected to the beam axis of two 
times the nuclear radius).

These pictures are the result of a transport
calculation for an impact parameter of $b$ = 7\,fm \cite{Hartnack:1900}.
The snapshots sketch the effect of pion shadowing by spectator matter
at different stages of the collision.
Those pions which are emitted in the early phase of the collision and
are detected around target rapidity  (i.e. at backward angles as indicated 
by the arrows in Fig.~\ref{figure-1}) will be shadowed by the projectile
spectator on one side  and therefore exhibit ``flow'' to the other side. 
In contrast, if pions freeze out at a late stage of the collision they 
will be shadowed (at target rapidity) by the target spectator  
which results in an ``antiflow''-like configuration.
Experimentally, we compare the number of pions emitted in the reaction 
plane to the side of the projectile spectator $N(\phi=0^{\circ})$ 
with the number of 
those pions emitted into the opposite direction $N(\phi=180^{\circ})$.
The azimuthal angle $\phi=0^{\circ}$ is defined by the projectile. 
$N_{\pi}(0^{\circ})$ 
refers to the angular range of $-45^{\circ}<\phi<45^{\circ}$, and 
$N_{\pi}(180^{\circ})$ to the angular range of $135^{\circ}<\phi<225^{\circ}$. 

Figure~\ref{figure-2} shows the ratio of these numbers for 
$\pi^+$ and $\pi^-$ mesons   
as a function of transverse momentum $p_T$ in two different 
rapidity regions and both for near-central and peripheral collisions.
In all four cases both $\pi^+$ and $\pi^-$ (open and full symbols) show a 
similar behavior ruling out Coulomb effects as the origin of the  
observed effect.
At mid rapidity the measured ratios are close to one both for near-central 
and peripheral collisions,  as expected for symmetric systems.
The deviations of about 10\% (for near-central collisions) 
reflect  the systematic error of the measurement
which is attributed mainly to the uncertainty in the determination of the 
reaction plane.

In near-central collisions (right panels in Fig.~\ref{figure-2}) 
the spectator fragments are small and hence 
shadowing effects are strongly reduced. Nevertheless, at target rapidities 
we observe that all pions - independent of their momentum -  
are emitted preferentially to the side of 
the target spectator (upper right panel of Fig.~\ref{figure-2}). 
The same asymmetry - which corresponds to pion ``flow'' -  was observed 
by the EOS collaboration at slightly higher incident 
energies \cite{Kintner:1997}.
Transport calculations have predicted a  
transition from pion ``antiflow'' to ``flow'' with decreasing impact parameter.
According to these calculations, pion ``flow'' in near-central collisions is 
a remainder of the flow of $\Delta$-resonances which decay into 
protons and pions \cite{Bass:1994}.

For peripheral collisions and at target rapidity 
 (upper left panel of Fig.~\ref{figure-2})
the ratio  $N_{\pi}(0^{\circ})/N(180^{\circ})$
decreases from about 1.2 at low p$_T$ values to about 0.5 at high p$_T$.
This behavior corresponds to a transition from pion number ``antiflow''
to ``flow'' with increasing transverse momentum. 
In earlier measurements it has been 
integrated over pion momentum  and hence, pion ``antiflow'' has been found 
since the pion yield is dominated by low-momentum pions\cite{Kintner:1997}.
According to transport calculations, ``antiflow'' is caused by  rescattering
of pions at the spectator matter in the late stage of the collision
\cite{Bass:1994}.
The transition from ``antiflow'' to ``flow'' as a function of 
transverse momentum as shown in the upper left panel of 
Fig.~\ref{figure-2} is a new 
observation which will be discussed in more details along with 
Fig.~\ref{figure-3}. 

A more detailed  picture of pion emission
is obtained by comparing the yield of pions emitted in plane to the 
yield of pions emitted perpendicular to the reaction plane. 
The latter pions are expected to be much less affected
by shadowing or rescattering by spectator matter and hence provide
a nearly undisturbed view onto the pion source. 
In order to visualize the effect of shadowing for different pion momenta
we normalize the in-plane pion spectra ($N_{\pi}(0^{\circ})$ 
and $N_{\pi}(180^{\circ})$) to the out-of-plane
spectra (N$_{\pi}(perp)$ = $(N_{\pi}(90^{\circ})+N_{\pi}(270^{\circ}))/2$).

Figure~\ref{figure-3} shows the ratios
$R_0=N_{\pi}(0^{\circ})/N_{\pi}(perp)$  
(``projectile side'', upper panel) and 
$R_{180}=N_{\pi}(180^{\circ})/N_{\pi}(perp)$  {``target side'', lower panel)
as a function of transverse momentum for peripheral collisions and 
target rapidities.
The ratios $R_{0,180}$ in Fig.~\ref{figure-3} do not exceed unity. This indicates
that the azimuthal asymmetry as shown in Fig.~\ref{figure-2} 
is not caused  by an enhanced
pion emission but rather by losses due to absorption or rescattering (which
result in ratios R inferior to unity). Figure~\ref{figure-3} 
allows to extract detailed
information on the emission time of pions as a function of their momentum.
At pion momenta around 0.4\,GeV/$c$, the upper left panel of 
Fig.~\ref{figure-2} exhibits
no asymmetry ($N_{\pi}(0^{\circ})/N_{\pi}(180^{\circ}) \approx 1$) whereas
Fig.~\ref{figure-3} clearly shows that
pion emission into the reaction plane is depleted ($R_{0,180} < 1$ both at the
projectile and target side). This effect is expected if pions are emitted
at about 13\,fm/$c$ when they are shadowed by both the
target and projectile spectator (see Fig.~\ref{figure-1}).
                                                       
Above momenta of 0.4\,GeV/$c$, the pion loss increases with increasing
momentum for pions emitted towards the projectile side (upper panel of 
Fig.~\ref{figure-3})
whereas the opposite trend is observed for pions emitted towards the
target side (lower panel of Fig.~\ref{figure-3}). 
This finding shows that high-momentum pions are correlated with early 
emission times (which are  even shorter than 13\,fm/$c$). 
In contrast, low-energy pions
predominantly freeze out at a later stage of the collision. This
information is based on the observation that pions with momenta below
0.3\,GeV/$c$ suffer from absorption or rescattering when emitted towards the
target side but remain undisturbed when emitted to the projectile side.
The depletion for low-energy pions (as shown in the lower panel 
Fig.~\ref{figure-3})
is less  pronounced than for high-energy pions (upper panel 
Fig.~\ref{figure-3}).
This effect indicates that low-energy pions freeze out over an extended
time span.
Again, $\pi^-$ and $\pi^+$ mesons behave very similarly
which demonstrates that the observed effects are not caused by Coulomb
interaction.

Before drawing conclusions on the time evolution of pion emission
we investigate another effect.
An anisotropy of the ratio $N_{\pi}(0^{\circ})/N_{\pi}(180^{\circ})$
as function of the pion transverse momentum may be caused by the
momentum dependence of the pion-nucleon cross section.
The same value of transverse momentum for pions emitted towards
the target and projectile remnants corresponds to different
relative momenta between pions and nucleons in the remnants.
We have performed calculations with a shadowing model using 
measured pion-nucleon scattering cross sections~\cite{PDG:2000} as
functions of the relative pion-nucleon momentum. 
The model allows to vary the correlation between pion energy and
emission time of the pion.
It turns out that the momentum-dependent cross section causes the 
ratio $N_{\pi}(0^{\circ})/N_{\pi}(180^{\circ})$ to decrease with 
increasing transverse momentum, qualitatively similar to the observation.
However, the calculated ratio $N_{\pi}(0^{\circ})/N_{\pi}(180^{\circ})$
remains well above unity for all transverse momenta up to 0.8\,GeV/$c$.
The observed reduction of the ratio 
$N_{\pi}(0^{\circ})/N_{\pi}(180^{\circ})$ to values well below 1 for
large transverse momenta can only be reproduced if hard pions are 
emitted prior to the instant of closest approach at 13~fm/$c$.

Our data show that in Au + Au collisions at 1\,GeV/nucleon
most of the high-energy pions freeze out within 13\,fm/$c$ after time 
zero.
Transport calculations predict a similar time scale for the emission
of high-energy pions ~\cite{Bass:1994,Li:1991}.  
According to  calculations, the nuclear density exceeds twice 
the saturation value in central Au+Au collisions at 1\,GeV/nucleon
within the first 15\,fm/$c$ \cite{Bass:1994}.
Therefore, the investigation 
of high-energy pions may open a new way to study the nuclear 
matter equation of state  at high baryonic densities.  

In summary, we have studied pion production in peripheral and near-central 
Au + Au collisions at 1\,GeV/nucleon as a function of pion transverse 
momentum, the azimuthal emission angle, and at different rapidities.    
In peripheral collisions at target rapidity, a reduced yield of
high-energy pions is observed at the projectile side.
This finding indicates that high-energy pions 
are shadowed by the incoming projectile spectator   
and, therefore, are emitted within the first 13\,fm/$c$ of the collision.
In contrast, low-energy pions observed at backwards angles 
are shadowed by the target spectator which means that they 
predominantly   freeze out in the late phase of the collision.

This work is supported by the Bundesministerium f{\"u}r Bildung und 
Wissenschaft, Forschung und Technologie under contract 06\,DA\,819 and by the
Gesellschaft f{\"u}r Schwer\-ionen\-forschung under contract DA\,OESK and 
by the Polish Committee of Scientific Research under contract No. 
2P3B11515. 


\begin{figure}
\begin{center}
\hspace{2.cm}\epsfig{file=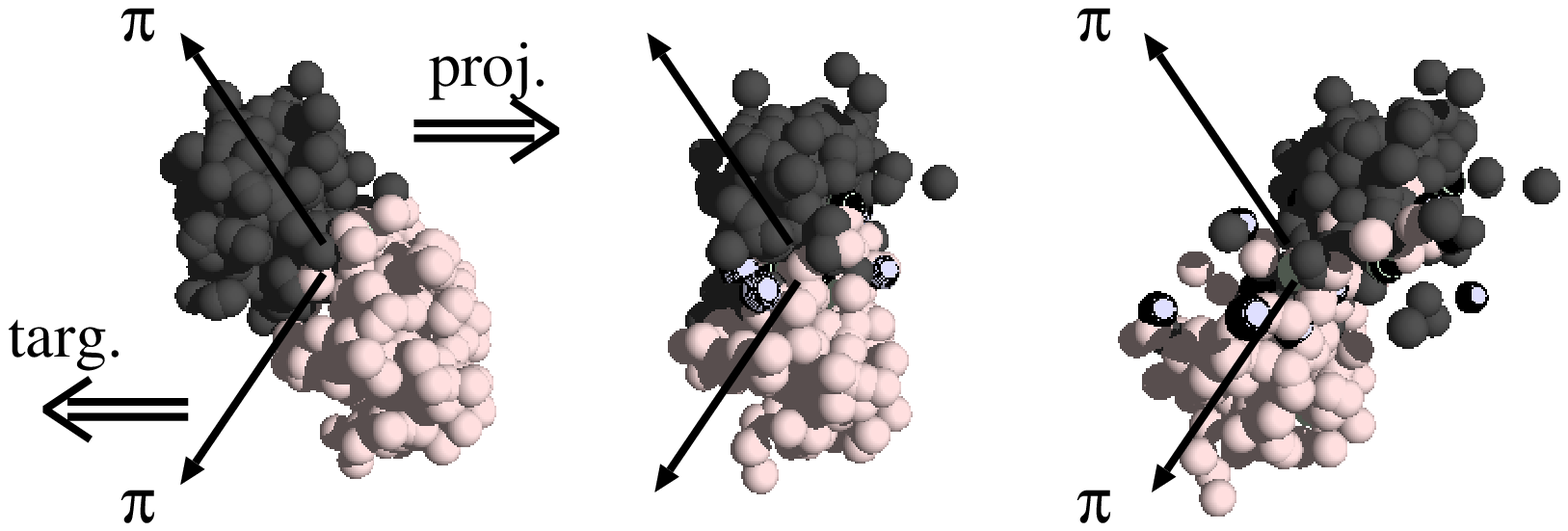,width=14cm}
\caption{ Sketch of an Au+Au collision at 1\,GeV/nucleon with an impact 
parameter of 7\,fm as calculated by a QMD transport code 
\protect\cite{Hartnack:1900}.
The snapshots are taken at 6.5\,fm/$c$ (left), 12.5\,fm/$c$ (middle), 
and 18.5\,fm/$c$ (right) after time zero (see text for definition
of time zero).
The arrows indicate the direction of the spectrometer at target rapidity.}
\label{figure-1}
\end{center}
\end{figure}

\begin{figure}
\epsfig{file=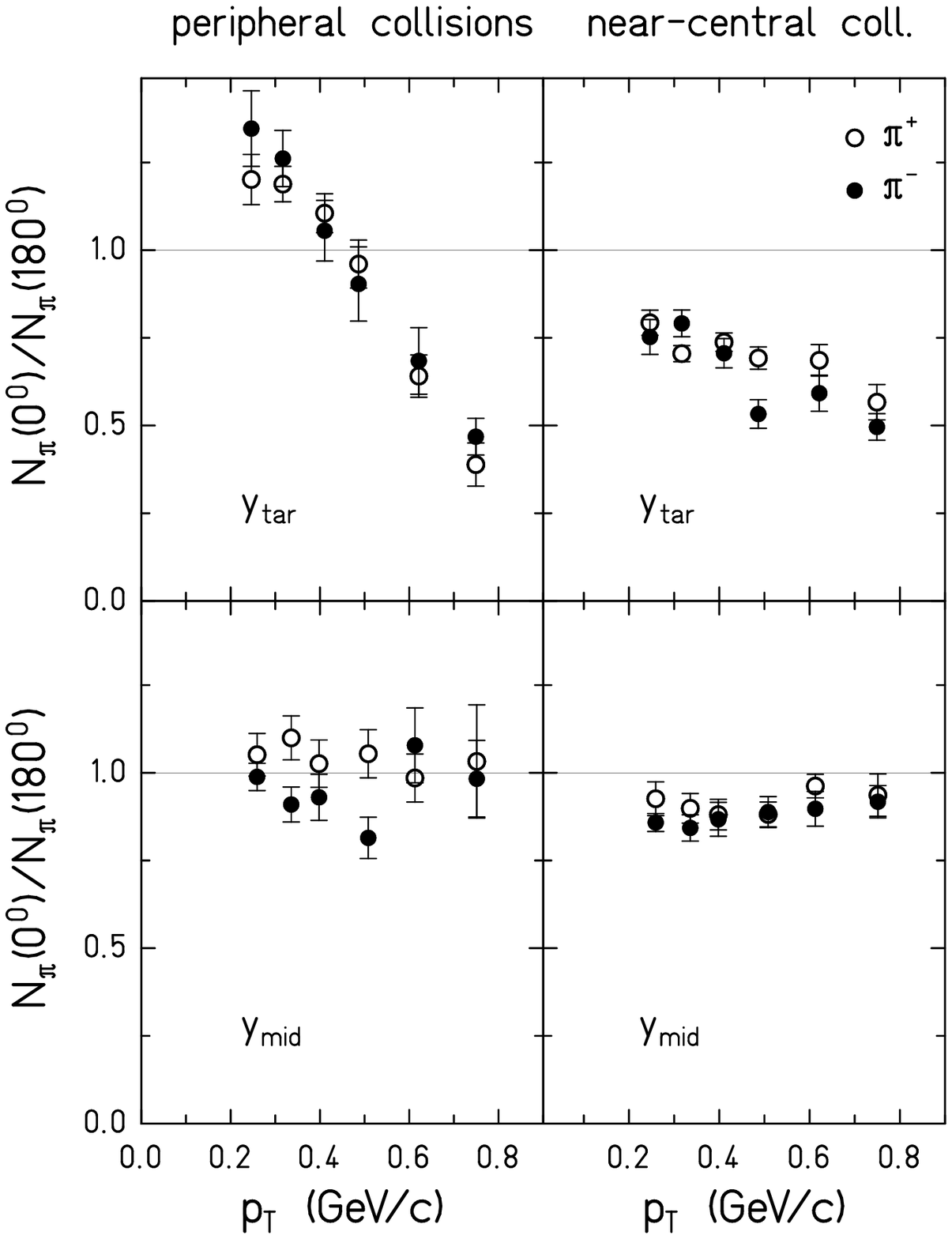,width=11cm}
\caption{ Pion number ratios $N_{\pi}(0^{\circ})/N_{\pi}(180^{\circ}$)
as a function of transverse momentum.
$N_{\pi}(0^{\circ}$) and $N_{\pi}(180^{\circ}$) denote the  number of 
pions emitted to the projectile and to the target side, respectively.
The spectra are measured in peripheral (left panel) 
and near-central (right panel) 
Au + Au collisions at 1\,GeV/nucleon and at two rapidity regions 
(the target rapidity region 
$0.01\cdot y_{beam}\leq y_{tar}\leq 0.10\cdot y_{beam}$ and the
mid-rapidity region
$0.44\cdot y_{beam}\leq y_{mid}\leq 0.56\cdot y_{beam}$).
Full (open) symbols refer to $\pi^-$ ($\pi^+$) emission. 
Only statistical errors are shown.}
\label{figure-2}
\end{figure}

\begin{figure}
\epsfig{file=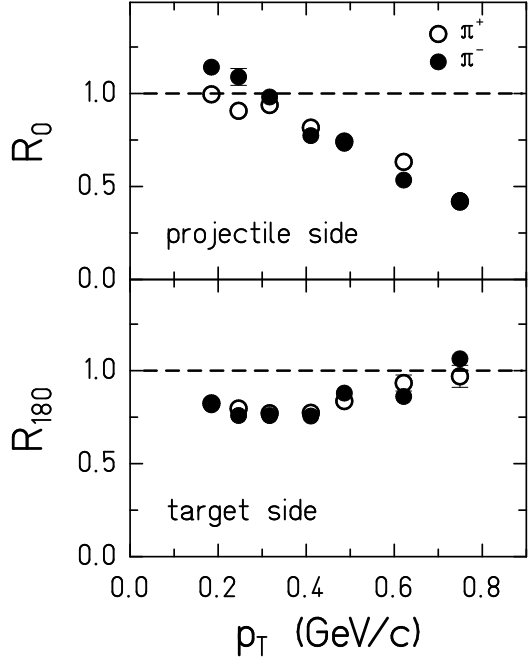,width=10cm}
\caption{ Pion ratios $R_0=N_{\pi}(0^{\circ})/N_{\pi}(perp)$ (upper panel)
and $R_{180}=N_{\pi}(180^{\circ})/N_{\pi}(perp)$ (lower panel) as a function 
of the transverse momentum. 
$N_{\pi}(0^{\circ})$ and $N_{\pi}(180^{\circ})$ 
denote the  number of pions emitted to the projectile and to the target side, 
respectively. $N_{\pi}(perp)$ is the number of pions emitted perpendicular to
the reaction plane (see text). The pions are measured  
in peripheral Au+Au collisions at 1\,GeV/nucleon at target rapidities
($0.01\cdot y_{beam}\leq y_{tar}\leq 0.10\cdot y_{beam}$).
Full (open) symbols refer to $\pi^-$ ($\pi^+$) emission.
Only statistical errors are shown.}
\label{figure-3}
\end{figure}
\end{document}